\documentclass[longbibliography,twocolumn,prl,aps,superscriptaddress,showpacs,amsmath,amssymb,floatfix]{revtex4-1}
\usepackage{graphicx}
\usepackage{amssymb}
\usepackage{amsmath}
\usepackage{epsfig}
\usepackage{color}
\usepackage{mathtools}
\usepackage[colorlinks,linkcolor=blue,anchorcolor=blue,citecolor=blue,urlcolor=blue]{hyperref}
\usepackage{physics}
\usepackage{bm}
\usepackage{tikz}
\usepackage{feynmf}
\usepackage[compat=1.1.0]{tikz-feynman}
\begin{document}

\title{Universality class and exact phase boundary in the superradiant phase transition}
\author{Wei-Feng Zhuang}
\affiliation{CAS Key Laboratory of Quantum Information, University of Science and Technology of China, Hefei, 230026, People’s Republic of China}
\author{Bin Geng}
\affiliation{CAS Key Laboratory of Quantum Information, University of Science and Technology of China, Hefei, 230026, People’s Republic of China}
\author{Hong-Gang Luo}
\affiliation{School of Physical Science and Technology, Lanzhou 730000, China}
\affiliation{Beijing Computational Science Research Center, Beijing 100084, China}
\author{Guang-Can Guo}
\affiliation{CAS Key Laboratory of Quantum Information, University of Science and Technology of China, Hefei, 230026, People’s Republic of China}
\affiliation {Synergetic Innovation Center of Quantum Information and Quantum Physics, University of Science and Technology of China, Hefei, Anhui 230026, China}
\affiliation{CAS Center For Excellence in Quantum Information and Quantum Physics, University of Science and Technology of China, Hefei, Anhui 230026, China}
\author{Ming Gong}
\email{gongm@ustc.edu.cn}
\affiliation{CAS Key Laboratory of Quantum Information, University of Science and Technology of China, Hefei, 230026, People’s Republic of China}
\affiliation {Synergetic Innovation Center of Quantum Information and Quantum Physics, University of Science and Technology of China, Hefei, Anhui 230026, China}
\affiliation{CAS Center For Excellence in Quantum Information and Quantum Physics, University of Science and Technology of China, Hefei, Anhui 230026, China}
\date{\today }

\begin{abstract}
The Dicke model and Rabi model can undergo phase transitions from the normal phase to the superradiant phase at the same boundary, which can be accurately determined 
using some approximated approaches. The underlying mechanism for this coincidence is still
unclear and the universality class of these two models is elusive. Here we prove this phase transition exactly using the path-integral approach based on the faithful 
Schwinger fermion representation, and give a unified phase boundary condition for these models. We demonstrate that at the phase boundary, the fluctuation of the bosonic field is 
vanished, thus it can be treated as a classical field, based on which a much simplified method to determine 
the phase boundary is developed. This explains why the approximated theories by treating the operators as classical variables can yield the exact boundary. 
We use this method to study several similar spin and boson models, showing its much wider applicability than the previously 
used approaches. Our results demonstrate that these phase transitions belong to 
the same universality by the classical Landau theory of phase transitions, which can be confirmed using the platforms in the recent experiments. 
\end{abstract}

\maketitle

The Dicke model has been studied for more than half a century \cite{dicke_coherence_1954,
	tavis_exact_1968,hepp1973superradiant, wang1973phase}. This model considers the coupling between 
$N$ identical atoms (or two-level systems) with a bosonic field, which can be written as
\begin{equation}
\mathcal{H} =  \omega b^\dagger b + \sum_{i=1}^N \frac{\Omega}{2} \sigma_i^z + {g \over \sqrt{N}} \sigma_i^x(b+b^\dagger).
	\label{eq-Dickemodel}
\end{equation}
Here $b$ is the annihilate operator for the bosonic field, $\sigma^x_i$, $\sigma^z_i$ are the Pauli operators for the $i$-th atom
and $N$ is the total number of atoms. This model undergoes a phase transition from a normal phase to a superradiant phase 
at $g_c^2 = \frac{\Omega \omega}{4} \coth {\beta \Omega \over 2}$ \cite{emary_chaos_2003,li_quantum_2006, Altland2012QuantumChaos}, where $\beta = 1/k_B T$, with 
$k_B$ is the Boltzmann constant and $T$ is the temperature. The phase transition can be obtained from the Holstein-Primakoff (HP) approximation \cite{emary_quantum_2003, holstein_field_1940}
and semiclassical approximation \cite{Bhaseen2012Dynamics, Grimsmo2013Dissipative}, in which negative or complex eigenvalues mark the ground state instability.
 It is challenging to be realized with atoms in radiation-plus-matter field due to not only the required large density, but also the no-go theorem \cite{rzazewski1975phase,
 bialynickibirula1979no-go,nataf2010no-go,chirolli2012drude}. However, it can be realized with ultracold atoms \cite{baumann_dicke_2010, baumann2011exploring, mottl2012roton, schmidt2014dynamical, landini_formation_2018, 
kroeze_spinor_2018, baden2014realization}, driven-dissipative quantum simulators \cite{damanet2019atom, dimer2007proposed} and spin-orbit coupled 
condensates in a trap \cite{Hammer2014Dicke} and electron gases in a cavity \cite{nataf2019rashba}.

Recently, the phase transition with only one atom  has attracted widespread attention \cite{hwang_quantum_2015, puebla2017probing,  liu_universal_2017,ashhab2013superradiance}. When $N=1$, Eq. \ref{eq-Dickemodel} is reduced to the exact solvable quantum Rabi model 
\cite{braak2011integrability,moroz2013solvability,batchelor2015integrability}. A great effort has been devoted in experiments trying to 
push the light-matter interaction strength $g$ from  the strong coupling regime (with $g$ larger than the dissipation rate) \cite{wallraff_strong_2004,
devoret2007circuit-qed, cristofolini_coupling_2012} to the ultrastrong coupling ($g \sim 0.1 \Omega$)  \cite{gunter_sub-cycle_2009,niemczyk_circuit_2010,peropadre_switchable_2010,
forn-diaz_observation_2010} and even the deep strong coupling regimes \cite{casanova_deep_2010}. This model has broad application in cold atoms \cite{ritsch2013cold},
trapped ions \cite{leibfried2003quantum,meekhof1996generation,haffner2008quantum,lv2018quantum},
quantum dots \cite{hanson2007spins}, cavity QED \cite{garziano2015multiphoton, law1996arbitrary} and superconducting circuits \cite{viehmann2011superradiant, peropadre_switchable_2010}.
It plays as a testing ground for strong coupling physics. Moreover, the calculation of its full spectra with the help of integrability is also of general interest \cite{Xie2017Review}.
It was shown \citep{hwang_quantum_2015,liu_universal_2017} that the phase transition is realized when $\frac{\omega}{\Omega} \to 0$  at $g_c$. In Ref. \cite{liu_universal_2017}, the critical exponent is shown to be in consistent with the Landau theory. The universal dynamics is also formulated using the Kibble-Zurek mechanism, which was originally established based on second-order phase 
transitions \cite{Zurek1985Cos, Zurek1996Cos}.

The phase transition in the Rabi model can be obtained using the simplest perturbation theory and the effective Hamiltonian approach by some truncation 
at $T=0$ \cite{hwang_quantum_2015,liu_universal_2017}. 
However, it is surprising that while approximations are involved, the predicted critical boundary is shown to be 
exact. This should not be regarded as some kind of coincidence, which is a long-standing unsolved problem in theory. Here 
we unveil the underlying origin based on the path-integral approach with Schwinger fermion representation. We demonstrate that the 
phase transitions in these two models belong to the same universality class by the Landau theory of phase transitions, from which the previous 
conclusions such as critical exponent and Kibble-Zurek dynamics in consistent with the mean-field theory will become straightforward. We give a unified boundary condition
\begin{equation}
	\omega = {(\eta g)^2 \over \Omega} \tanh \frac{\beta \Omega}{2}, \quad    \frac{\omega}{N\Omega} \to 0.
    \label{eq-unified}
\end{equation}
Physically, it means a classical phase transition since the quantization of the bosonic field is vanishing. The parameter $\eta$ in both models accounts for the effect of rotating 
wave approximation, in which $\eta = 1$ for the presence of it,
and $\eta = 2$ for its absence. At this point, the fluctuation of the bosonic field is negligible, based on which we derive a much simpler method to study the 
phase transition in some similar models with interaction between boson fields and atoms, showing that the same phase transition can happen in models with nonidentical atoms, 
Hubbard interaction and nonlinearity, all of which belong to the Landau paradigm of phase transition. This method is demonstrated to have much broader applicability than the 
previous approximated approaches. 

We implement Eq. \ref{eq-Dickemodel} using the path-integral approach developed by Popov \cite{popov_behavior_1982,alcalde_path_2011,popov1988functional,
popov1991functional, alcalde_path_2011}, in which the spins are represented by the Schwinger fermions \cite{baskaran1987resonating,wen1996theory}
\begin{equation}
	\sigma_i^+  = \frac{\sigma_i^x +i \sigma_i^y}{2} =  \alpha_i^\dagger \beta_i, \quad \sigma_i^z = \alpha_i^\dagger \alpha_i - \beta_i^\dagger \beta_i.
\end{equation}
Here $\alpha_i$ and $\beta_i$ are fermion operators, and the number of fermions $\sum_{i} \alpha_i^\dagger \alpha_i + \beta_i^\dagger \beta_i = N_F$ is constrained by the number of spins $N_F= N$. 
This is a faithful representation since the Hilbert spaces in these two theories are the same. In the few-spin models, we can even prove that the partition function of the
Hamiltonian in these two representations are exactly the same. This is different from the HP method, in which the spin and boson have different Hilbert 
spaces. The partition function reads as
\begin{equation}
	Z = \Tr \exp(-\beta \mathcal{H} ) = i ^N \Tr \exp (-\beta \mathcal{H}_F - {i \pi \over 2} N_F),
	\label{partition_fun}
\end{equation}
where $N_F$ is the constraint defined above. In this new representation, we can take the constraint into account  and write the partition function in terms of these fermions as
following
\begin{equation}
Z= i^N \int\mathcal{D}\bar{\alpha}\mathcal{D}\alpha\mathcal{D}\bar{\beta}\mathcal{D}\beta\mathcal{D}\bar{b}\mathcal{D}b\text{e}^{-S},
\label{eq-partition}
\end{equation} 
where $S=\int\big(\bar{b}\frac{\partial}{\partial\tau}b+\sum_{i=1}^{N}\bar{\alpha}_{i}\frac{\partial}{\partial\tau}\alpha_{i}+\bar{\beta}_{i}\frac{\partial}{\partial\tau}\beta_{i} + \mathcal{H}\big)d\tau$.
We first make a rotating wave approximation to Eq. \ref{eq-Dickemodel}, which corresponds to the Jaynes-Cummings model. Via the fermion coherent representation we have
\begin{equation}
	\mathcal{H} = \omega\bar{b}b+\sum_{i}\frac{\Omega}{2}(\bar{\alpha}_{i}\alpha_{i}-\bar{\beta}_{i}\beta_{i})+ {g \over \sqrt{N}} (\bar{\alpha}_{i}\beta_{i} b +\bar{\beta}_{i}\alpha_{i} \bar{b}) + 
	{i\pi \over 2\beta} N_F.
\end{equation}
The trace in Eq. \ref{partition_fun} is carried out over different $N_F$ spaces of $\mathcal{H}$, in which only the state with $N_F=N$ is physical, while all the other modes are canceled exactly \cite{popov1988functional, alcalde_path_2011, alcalde_path_2011}. We solve the above  model based on Fourier transformation $b(\tau)=\sum_{n}b_{n}e^{i\omega_{n}\tau}$ and 
$\psi_{i}(\tau)=\sum_{q}\psi_{i}(q)e^{i\omega_{q}\tau}$, 
with $\psi_i$ for fields $\alpha_i$ and $\beta_i$, where $\omega_n=2n\pi/\beta$, $\omega_q=(2q+1)\pi/\beta$ ($n, q  \in \mathbb{Z}$) are Matsubara frequencies for 
bosons and fermions. The total action is decoupled into two parts $S = S_0 + S_\text{int}$, where $S_{0} = \sum_{k,q}\bar{\psi}_{k}(q)G_{0}^{-1}(q)\psi_{k}(q)$, with 
$\psi_k(q)=(\alpha_k(q), \beta_k(q))^T$, and  
\begin{equation}
	G_0(q) = \begin{pmatrix}
				\mathcal{G}^+_q   & 0 \\ 
					0   & \mathcal{G}^-_q 
			\end{pmatrix}, \quad \mathcal{G}^{\pm }_q= \frac{1}{\beta(i\omega_q+i\frac{\pi}{2\beta}\pm \frac{\Omega}{2})}. 
			\label{eq-G0}
\end{equation}
The interaction term can be written as $S_{\text{int}} =\sum_{k}\sum_{q,q^{\prime}}\bar{\psi}_{k}(q)\Sigma(q-q^{\prime})\psi_{k}(q^{\prime})$, where
\begin{eqnarray}
	\Sigma(q-q^{\prime}) =\frac{g \beta}{\sqrt{N }}\begin{pmatrix}0 & b_{q-q^{\prime}} \\
		\bar{b}_{q^{\prime}-q} & 0
\end{pmatrix}.
\label{eq-Sigmaint}
\end{eqnarray}

We see that for the fermion fields, the interacting term is in a quadratic form; while for the bosonic field by treating the fermion field as a Grassmannian constant, the 
interacting term is just a linear displacement of the bosonic field. We take advantage of this feature and integrate out of the fermion fields $\psi_i$,
leaving only the bosonic field in the following form $Z =\int\mathcal{D}\bar{b}\mathcal{D}be^{-S_{\text{eff}}[\bar{b},b]}$, where 
\begin{eqnarray}
	S_{\text{eff}}[\bar{b},b] =\sum_{n}\beta(i\omega_{n}+\omega)\bar{b}_{n}b_{n}-N\tr \ln G^{-1},
	\label{eq-eff_action}
\end{eqnarray}
with $G^{-1}=G_{0}^{-1}+\Sigma$. The previous literature tries to solve the above model from the saddle point solution of 
$S_\text{eff}$ \cite{alcalde_path_2011, popov_behavior_1982,alcalde_path_2011,popov1988functional,
popov1991functional} and its fluctuation around this point. We choose a different strategy by expanding the solution 
to infinite orders via Taylor expansion of the bosonic field. In the second term of $S_{\text{eff}}$, we utilize $-\tr \ln G^{-1}= -\tr\ln G_0^{-1} + 
\tr \sum_{ m \ge 1}\frac{1}{2 m}(G_{0}\Sigma)^{2m}$, which can be represented using the following Feynman diagrams
\begin{eqnarray} 
& &	-\tr \ln G^{-1}  =  -\tr \ln G_0^{-1}  +  \nonumber	\\
& &  
\begin{tikzpicture}[baseline=-3]
\begin{feynman} 
\vertex (o) ;
\vertex [left=20pt of o] (a1) ;
\vertex [right=20pt of o] (a2) ;
\vertex [right=20pt of a2] (i2) ;
\vertex [left=20pt of a1] (i1);
\diagram* {
	{
		(a2) --[edges=fermion,out=90,in=90,looseness=1.6,edge label'=$\mathcal{G}^+_q$,arrow size=1.2pt] (a1),
		(a1) --[edges=fermion,out=-90,in=-90,looseness=1.6,edge
		label'=$\mathcal{G}^-_{q-n}$,arrow size=1.2pt] (a2),
		(a1) --[edges=charged boson,edge label'={$n$},arrow size=1.2pt] (i1), 
		(i2) --[edges=charged boson,edge label'={$n$},arrow size=1.2pt] (a2) 
	}
};  
\end{feynman}
\end{tikzpicture}   +
\frac{1}{2} 
\begin{tikzpicture}[baseline=-3]
\begin{feynman}
\vertex (o);
\vertex [above=25pt of o] (a1);
\vertex [left=25pt of o] (a2);  
\vertex [below=25pt of o]  (a3);
\vertex [right=25pt of o] (a4);
\vertex [above=22pt of a1] (i1);
\vertex [left=22pt of a2] (i2);
\vertex [below=22pt of a3] (i3);
\vertex [right=22pt of a4] (i4);
\diagram* 
{    
	{(a1) --[edges=fermion,out=180,in=90,edge label'=$\mathcal{G}^+_q$,arrow size=1.2pt] (a2)},
	{(a2) --[edges=fermion,out=-90,in=180,edge label'=$\mathcal{G}^-_{q-n_1}$,arrow size=1.2pt] (a3)},
	{(a3) --[edges=fermion,out=0,in=-90,edge label'=$\mathcal{G}^+_{q-n_1+n_2}$,arrow size=1.2pt] (a4)},	
	{(a4) --[edges=fermion,out=90,in=0,arrow size=1.2pt] (a1)},
	{(i1) --[edges=charged boson,edge label={$n_1-n_2+n_3$},arrow size=1.2pt] (a1)},
	{(a2) --[edges=charged boson,edge label'={$n_1$},arrow size=1.2pt] (i2)},
	{(i3) --[edges=charged boson,edge label={$n_2$},arrow size=1.2pt] (a3)},
	{(a4) --[edges=charged boson,edge label'={$n_3$},arrow size=1.2pt] (i4)},	
};
\end{feynman}
\end{tikzpicture} + \cdots  
\nonumber  \\
&&  + 
\frac{1}{m} 
\begin{tikzpicture}[baseline=-3]
\begin{feynman}
\vertex (o)  ;
\vertex [above=30pt of o] (a1) ;
\vertex [above left=9.3pt and 28.5pt of o] (a2) ;
\vertex [below left=24.3pt and 17.6pt of o] (a3) ;
\vertex [below right=24.3pt and 17.6pt of o] (a4)  ;
\vertex [above right=9.3pt and  28.5pt of o] (a5) ;
\vertex [above =28pt of a1] (i1) ;
\vertex [above left=8.6pt and 26.6pt of a2] (i2) ;
\vertex [below left=22.6pt and 16.4pt of a3]  (i3) ;
\vertex [below right=22.6pt and 16.4pt of a4] (i4) ;
\vertex [above right=8.6pt and 26.6pt  of a5] (i5) ;

\diagram* {
	{(a1) --[edges=fermion,out=180,in=72,edge label'=$ \mathcal{G}_q^+$,arrow size=1.2pt]  (a2)},
	{(a2) --[edges=fermion,out=252,in=144,edge label'=$\mathcal{G}_{q-n_1}^-$,arrow size=1.2pt] (a3)},
	{(a3) --[edges=fermion,out=324,in=216,edge label'=$\mathcal{G}_{q-n_1+n_2}^+$,arrow size=1.2pt] (a4)},
	{(a4) --[edges=ghost,out=36,in=288,edge label'=$\cdots$,arrow size=1.2pt] (a5)},
	{(a5) --[edges=fermion,out=108,in=0,arrow size=1.2pt] (a1)},
	{(i1) --[edges=charged boson,edge label={$n_1-n_2 \cdots +n_{2m-1}$},arrow size=1.2pt] (a1)},
	{(i1) --[edges=charged boson,arrow size=1.2pt] (a1)},	
	{(a2) --[edges=charged boson,edge label={$n_1$},arrow size=1.2pt] (i2)},
	{(i3) --[edges=charged boson,edge label={$n_2$},arrow size=1.2pt] (a3)},
	{(a4) --[edges=charged boson,edge label={$n_3$},arrow size=1.2pt] (i4)},
	{(a5) --[edges=charged boson,edge label'={$n_{2m-1}$},arrow size=1.2pt] (i5)}
};
\end{feynman} 
\end{tikzpicture}+\cdots.
\label{eq-expansion}
\end{eqnarray}

In these diagrams, the bosonic field can be written as 
\begin{equation}
	\mathcal{V}_{\{n_i\}}^{(2m)} = \sum_{\{ n_i \}}\chi_{\{n_i\}}^{(2m)} b_{n_1} \bar{b}_{n_2} b_{n_3} \bar{b}_{n_4} \cdots b_{n_{2m-1}} \bar{b}_{k_{2m}},
\end{equation}
for $k_{2m} =  \sum_{i=1}^{2m-1} (-1)^{i+1} n_i$, in which the Matsubara summation of $\omega_q $ is performed. The leading term yields
\begin{equation}
S_{\rm {eff}}^{(2)}=\sum_{n} \bigg(i\omega_{n}+\omega-\frac{g^{2}}{i\omega_{n}+\Omega}\cdot\tanh\frac{\beta\Omega}{2}\bigg)|b_n|^2.
\end{equation}
The real part of this expression, which should be positive for all modes for the normal phase, has been used to determine the superradiant phase transition 
in the Dicke model \cite{popov_behavior_1982,alcalde_path_2011,popov1988functional, popov1991functional}. It is given by the mode $b_0$ by 
Eq. \ref{eq-unified} with $\eta = 1$. However, whether the phase transition occurs or not also depends critically on the higher-order terms \cite{firstorder}. 
The next leading term in $S_{\text{eff}}$ is $\chi_0^{(4)}  |b_0|^4$, where 
\begin{equation}
 \chi_{0}^{(4)}=\frac{g \beta }{2N}\bigg(\frac{g}{\Omega}\bigg)^{3} \bigg( 2\tanh \frac{\beta \Omega}{2} - 
	\beta\Omega \sech^2 \frac{\beta\Omega}{2}  \bigg).
\end{equation}
We see that $\chi_0^{(4)}$ is always a positive number. When $n_i$ are different and when all the singular points are first order, we can 
perform  the Matsubara summation of $\omega_q$ via the residue theorem and obtain
\begin{equation}
\chi_{\{n_i\}}^{(4)}=\frac{g^4 \beta \tanh \frac{\beta\Omega}{2}  \left(i \omega_{n_1+n_3}+2 \Omega \right)}{ 2 N (i \omega_{n_1-n_2+n_3}+\Omega)  \prod_{i=1}^{3} ( i \omega_{n_i} + \Omega)}.
\end{equation}
This is a complex value, with its real part to be either positive or negative, depending strongly on the values of $n_i$. We find the following upper bound
\begin{equation}
	|\chi_{\{ n_i \}}^{(4)}| \le \frac{ \sqrt{6} g \beta }{N} \bigg( \frac{g}{\Omega} \bigg)^{3} \tanh \frac{\beta\Omega}{2}
	\le \frac{ \sqrt{6} g \beta }{N} \bigg( \frac{g}{\Omega} \bigg)^{3}.
\end{equation}
This result can be generalized to arbitrary orders. For the $2m$-th term, we can read from the Feynman diagram as 
$\chi^{(2m)}_{\{n_i\}} =  \sum_{q} {g^{2m} \over N^{m-1} } \mathcal{G}_q^+ \mathcal{G}_{q-n_1}^- \mathcal{G}_{q-n_1 + n_2}^+ \cdots \mathcal{G}_{k_{2m}}^-$.
We noticed that for the higher-order terms, this summation is absolute convergent during the Matsubara summation of $\omega_q$. Let us choose a sufficient large $|q| < K_m /2$, 
then using $|\mathcal{G}_q^\pm| \le 2/\beta |\Omega|$, we have
\begin{equation}
	|\chi^{(2m)}_{\{n_i\}}| < \frac{K_m}{N^{m-1}} \bigg(\frac{g}{\Omega} \bigg)^{2m}.
	\label{eq-Km}
\end{equation}
This upper bound is independent of $n_i$. The value of $\chi^{(2m)}_{\{n_i\}}$ can be calculated using the residue theorem, in which the high order 
singular points need to be treated carefully. However, this upper bound does not reply on the feature of the singular points, thus has much wider 
applicability. In general, the large $m$ is, the smaller $K_m$ will be for the reason of fast convergence of Matsubara summation of $\omega_q$. 

This estimation can also be applied to the full Dicke model. In this case, the self-energy $\Sigma$ should be changed accordingly 
by setting its off-diagonal component $b_{q-q'}$ to $(b_{q-q'} + \bar{b}_{q' -q})$ in Eq. \ref{eq-Sigmaint}. However, the propagator $\mathcal{G}_q^\pm$ is unchanged. Thus the above 
upper bound is still applicable. We have 
\begin{equation}
	S_\text{eff}^{(2)} =\sum_{n} (i\omega_{n}+\omega) \bar{b}_n b_n - \frac{g^{2} \tanh\frac{\beta\Omega}{2}}{i\omega_{n}+\Omega}(b_n + \bar{b}_{-n})^2,
\end{equation}
which has the same symmetry --- $U(1)$ in the Jaynes-Cummings model with $\eta =1$ and $\mathbb{Z}_2$ in the Dicke and Rabi models with $\eta =2$ --- as the original Hamiltonian. 
The above action yields the boundary in Eq. \ref{eq-unified}. 

This expression can lead to some remarkable results. When $N\rightarrow \infty$, 
the higher-order terms will approach zero, thus only the leading term $S_\text{eff}^{(2)}$ is important. In the few particle case, this limit can be
reached via $\frac{g^2}{\Omega} \rightarrow 0$, which is equivalent to $\frac{\omega}{\Omega} \rightarrow 0$ using the solution of $g_c^2 \sim \Omega \omega$ at 
the boundary. This result yields the constraint in Eq. \ref{eq-unified} in a unified form, which corresponds to the classical limit of the bosonic field.
It has some immediate consequences. In this condition,
the higher-order terms of the bosonic field will disappear, leaving only the term $b_0$ to be important. Thus 
we only  need to treat the field as a classical variable, for which reason the classical treatment is accurate for the phase transition. Let us assume $b \rightarrow b_0$ 
and $b^\dagger \rightarrow b_0^*$, then
\begin{equation}
	\mathcal{H} =  \omega |b_0|^2 + {\Omega \over 2} \sum_i \sigma^z_i  +  {g \over \sqrt{N}}\sum_{i}^N( \sigma_i^\dagger  b_0 + b^* \sigma_i^-).
\end{equation}
The $N$ two-level atoms are now independent. We can calculate the free energy of the above model at finite temperature from 
$Z = e^{-\beta F} = \Tr(e^{-\beta \mathcal{H}})$, which yields $F = \omega |b_0|^2 - \frac{N}{\beta} \ln {\cosh(\beta E) \over 2}$ with $E = \sqrt{\Omega^2/4 + g^2 |b_0|^2/N}$. 
Around $b_0 \sim 0$, we have
\begin{equation}
	F = F_0 + |b_0|^2 \bigg(\omega - {g^2 \tanh {\beta \Omega \over 2} \over \Omega} \bigg) + \sum_{n \ge 2} F_{2n} |b_0|^{2n},
	\label{eq-F0F1F2}
\end{equation}
where $F_0 = -\frac{N}{\beta} \ln (2\cosh {\beta \Omega \over 2})$. This result yields Eq. \ref{eq-unified}. Here, $b_0$ is a classical variable, thus it 
forbids the superposition of two different states in the spontaneous symmetry breaking phase. The higher-order terms $F_{2n}$ are ignored in the previous literature for the phase 
transition \cite{hwang_quantum_2015, puebla2017probing,  liu_universal_2017,ashhab2013superradiance}, in which the proof may suffer from loopholes. 
We can prove that the higher-order terms will scales as $(g^2/(N\Omega^2))^n$, which is in accord with Eq. \ref{eq-Km}. 
However, their signs are alternating. When $\beta \Omega \gg 1$, we have
\begin{eqnarray}
	F_4 \rightarrow {g^4 \over N\Omega^3}, \quad F_6 \rightarrow - {2g^6 \over N^2 \Omega^5}, \quad 
	F_8 \rightarrow {5 g^8 \over N^3 \Omega^7}.
\end{eqnarray}
In general, $F_{2n} \sim (-1)^n \frac{\Omega }{N^{n-1}}(\frac{g}{ \Omega})^{2n} $. 
The negative coefficients of the higher-order terms may lead to failure of Landau theory of phase transitions (e.g., see the first-order phase transitions by Landau theory in Ref. 
\onlinecite{firstorder}). To ensure of exact second-order phase transition, one requires that all these 
coefficients are vanished, which can be achieved only when the constraint in Eq. \ref{eq-unified} is satisfied, leaving only the leading term $S_\text{eff}^{(2)}$ for instability. In this 
sense, at the critical point, the fluctuation of the bosonic field is negligible, making our conclusion to be exact. This justifies why even the simplest approximations in the previous 
literature can yield the accurate phase boundary. It also means that the phase transition is exactly described by the 
Landau theory with the number of photon as $ \langle b^{\dagger}b\rangle\sim|g-g_{c}|^{-\alpha}$, where $\alpha = 1/2$ is the same as the 
mean-field theory \cite{hwang_quantum_2015,liu_universal_2017,hwang_quantum_2015,kirton2019introduction,nagy2011critical,nagy2016critical}.

\begin{figure}
	\centering
	\includegraphics[width=0.45\textwidth]{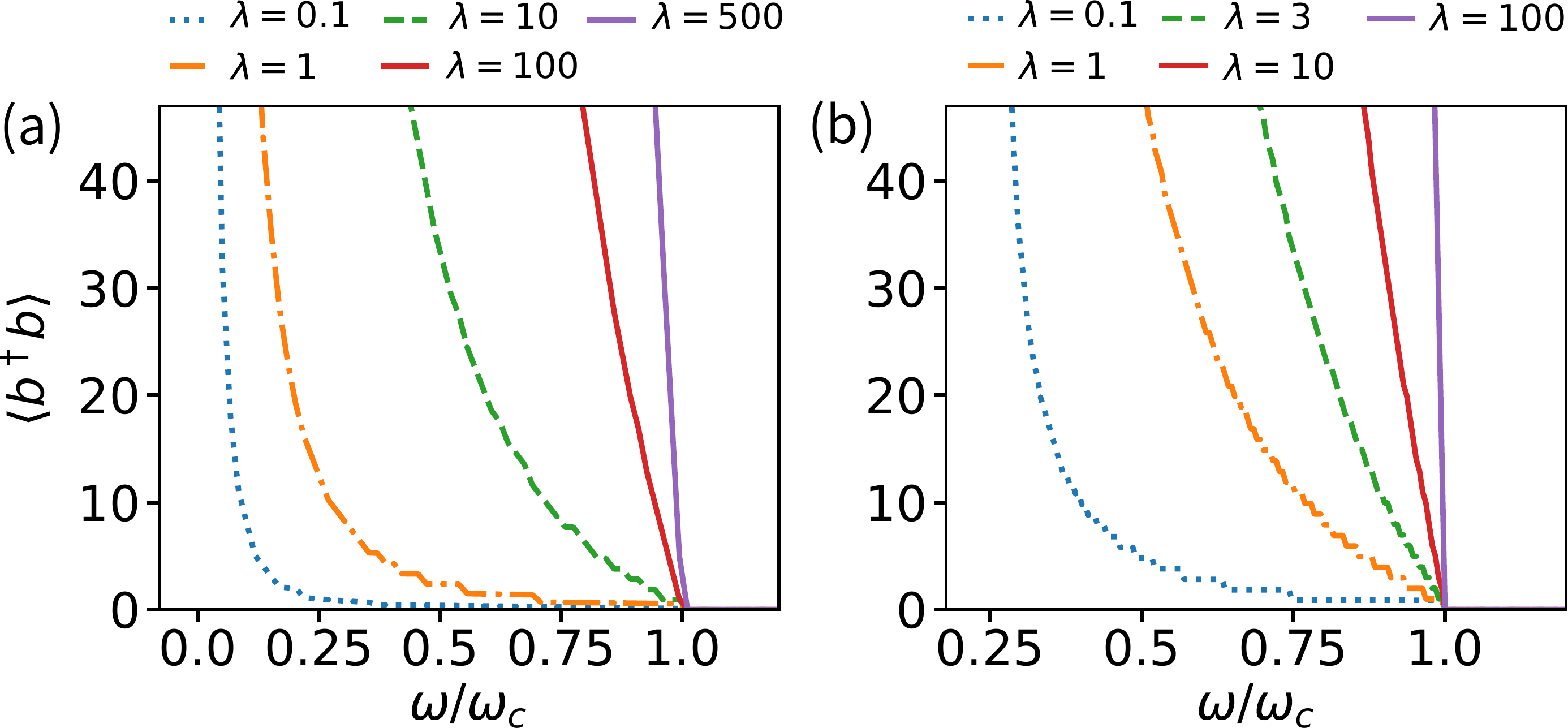}
	\caption{Phase transitions for model (I) in (a) and model (III) in (b) based on exact diagonalization method at $T= 0$. In (a), we choose 
	$N=3$, $(\tilde g_1 ,\tilde g_2, \tilde g_3) = (12, 1, 100)$, and $(\tilde \Omega_1, \tilde \Omega_2, \tilde \Omega_3) = (9,50,110)$.  Different lines 
	are plotted with $g_i = \sqrt{\lambda} \tilde{g}_i$ and $\Omega_i = \lambda \tilde{\Omega}_i$, which yield 
	$\omega_c = 35.6$ from Eq. \ref{eq-boundaryI}. In (b), we choose $N=1$, $\kappa = 0.5$, $\tilde{g} = 12$ and  $\tilde{\Omega} = 100$, with 
	$\omega_c = 1.94$ from Eq. \ref{eq-boundaryIII} using $\lambda$ defined in the same way as (a). }
	\label{fig-fig1} 
\end{figure}

Our result is useful to understand the phase transitions in the other models with spin and boson interaction for second-order phase transition, in which the HP and the 
effective Hamiltonian approaches are failed. We discuss several models (I) - (III), which have been justified by exact numerical method with high accuracy (Fig. \ref{fig-fig1}). 
This approach applies to physics even at finite temperature. 

(I) Phase transition in the inhomogeneous model. This model reads as
\begin{equation}
	\mathcal{H}=\omega b^\dagger b + \sum_i \Omega_i \sigma^z_i + \sum_i (\frac{g_i}{\sqrt{N}} \sigma_i^\dagger b + \text{h.c.}),
\end{equation}
for nonidentical atoms interact with a common field. The dynamics in this model has been studied in [\onlinecite{tsyplyatyev2009dynamics,strater2012nonequilibrum,zou2014implementation}]. The phase transition happens at 
\begin{equation}
	\omega=   \frac{1}{N} \sum_{i=1}^N   \frac{g_i^2}{ \Omega_i} \tanh \frac{\beta \Omega_i}{2}, \quad  \frac{1}{N} \sum_i\frac{g_i^2}{\Omega_i} \to 0.
	\label{eq-boundaryI}
\end{equation}
This expression is in consistent with the result in Ref. \cite{hioe1973phase} with	inhomogeneous interaction. We confirm this phase transition in 
Fig. \ref{fig-fig1} (a), in which a divergent of $\langle b^\dagger b\rangle$ is expected from $S_\text{eff}^{(2)}$ across the phase boundary 
due to the vanished higher-order terms \cite{kirton2019introduction}.

(II) Anti-rotating term and Hubbard interaction. In this case, we consider the anisotropic interaction of the form of 
$(g_{1} \sigma_{i}^{+} +g_{2} \sigma_{i}) b/\sqrt{N} + \text{h.c.}$ and Hubbard interaction of $U n(n-1)$, with $n = b^\dagger b$. It is frequently termed as anisotropic Rabi model
when $g_1 \ne g_2$  \cite{Xie2014Anisotropic}. In this case we find the energy level spacing mediated by this term is $\sqrt{\Omega^2/4 + |g_1 b_0 + g_2 b_0^*|^2 /N}$, which preserves 
the $\mathbb{Z}_2$ symmetry. The Hubbard term $U$ is unimportant for the phase transition. We have phase transition at 
\begin{equation}
	\omega = {(g_1 + g_2 )^2  \over \Omega} \tanh {\beta \Omega \over 2}, \quad {\omega \over N\Omega} \rightarrow 0.
	\label{eq-boundaryII}
\end{equation}
This condition has been shown in literature \cite{hwang_quantum_2015,liu_universal_2017}, and it can be manifested much more straightforward 
in this work. Thus we have $\eta = 2$ in Eq. \ref{eq-unified} when all $g_i=g$.

(III) Nonlinearity effect. It is inevitable that the higher-order correction by the bosonic field can slightly modify the energy level spacing of the atoms \cite{dimer2007proposed}. 
We mimic this effect using the model 
$\mathcal{H} =  \omega b^\dagger b + \sum_i (\Omega/2 + \kappa b^\dagger b) \sigma^z_i + g/\sqrt{N} \sum_i (b^\dagger \sigma_i + \text{h.c.})$, where the term $\kappa$ maybe introduced
via the higher-order perturbation theory. This model can not be solved by the HP method for the reason of nonlinear interaction between them. We find the phase transition 
happens at 
\begin{equation}
\omega = { g^2 + \kappa N  \Omega \over \Omega} \tanh {\beta \Omega  \over 2}, \quad {\omega \over N \Omega} \rightarrow 0.
\label{eq-boundaryIII}
\end{equation}
We confirm this phase transition in Fig. \ref{fig-fig1} (b). 
This result will have some interesting predictions. When $\kappa$ is independent of $N$, it is relevant, and the phase transition is forbidden in the thermodynamic limit.
When $\kappa = \kappa_0/N$, which is most likely to happen since the bosonic field is proportional to $1/\sqrt{N}$, we find that this phase transition is still presented. 
However, when $\kappa \propto \kappa_0/N^{\gamma}$, where $\gamma > 1$, this nonlinear effect is irrelevant in the thermodynamic limit, which will not influence the 
phase boundary. Thus $\gamma = 1$ is marginal. This result means that the Dicke phase transition can still happen even taken the nonlinear correction into account.

To conclude, this work is stimulated by the coincident phase boundary in the Dicke and quantum Rabi models, which is exact though derived by some approximated 
approaches. We explore the underlying origin using the path-integral approah and give a unified boundary condition for these two models, at which the fluctuation of the 
bosonic field is vanished. In this limit, we can treat the bosonic field as a classical variable, which has much broader applicability than all the above approximated 
approaches in the determination of phase boundaries in some of the spin and boson interacting models. All these phase transitions belong to the classical Landau theory 
of phase transition, thus the critical exponent and the associated universal dynamics should be the same as that from the mean-field theory, which can be confirmed using 
cold atoms, trapped ions and superconducting circuits. 

\textit{Acknowledgements.} This work is supported by the National Natural Science Foundation of China (No. 11774328 and No. 11834005) and the National Key Research and Development Program of China (No. 2016YFA0301700).

\bibliography{ref}

\begin{thebibliography}{68}
\bibitem [{fir()}]{firstorder}%
  \BibitemOpen
  \href@noop {} {}\bibinfo {note} {First order phase transition can be realized
  by Landau theory in the following model $f(T) = f_0 + \alpha (T-T_c) m^2 +
  \beta m^4 + \gamma m^4$, where $\beta < 0$ and $\alpha >0$, $\gamma >
  0$.}\BibitemShut {Stop}%
\end{thebibliography}

\end{document}